\documentclass[times,10pt,twocolumn]{article}
\usepackage{latex8}
\usepackage[]{pslatex}     

\usepackage[]{graphicx}
\usepackage[]{subfigure}  
\usepackage[]{dcolumn}    
\usepackage[]{amsmath}    

\newcounter{bean}
\newenvironment{tightenum}%
{
\begin{list}%
{\arabic{bean}}{\usecounter{bean}
    \setlength{\topsep}{0in}
    \setlength{\partopsep}{0in}
    \setlength{\itemsep}{0in}
    \setlength{\parsep}{0.05in}
    \setlength{\leftmargin}{1.0em}
    \setlength{\rightmargin}{0in}
    \setlength{\itemindent}{0in}
}
}%
{ \end{list}}

\begin{document}

\clubpenalty=9999
\widowpenalty=9999
\def\>{\rangle}
\def\be{\begin{equation}}
\def\ee{\end{equation}}
\newcommand{\horizbar}{\rule{\linewidth}{.5mm}}
\newcommand{\app}[1]{{\sc #1}}
\newcommand{\bra}[1]{\langle#1|}
\newcommand{\ket}[1]{{|}#1{\rangle}}
\newcommand{\logical}[1]{\overline{#1}}
\newcommand{\xhat}{\hat{x}}
\newcommand{\yhat}{\hat{y}}
\newcommand{\zhat}{\hat{z}}
\newcommand{\gt}[1]{\textsc{#1}}
\newcommand{\cnot}{\gt{cnot}}
\newcommand{\swap}{\gt{swap}}
\newcommand{\xor}{\gt{xor}}
\newcommand{\rootswap}{\sqrt\gt{swap}}
\newcommand{\eccode}[3]{$[\![#1,#2,#3]\!]\;$}
\newcommand{\ecc}{\eccode{7}{1}{3}}
\renewcommand{\em}{\it}
\newcommand{\BigO}[1]{${\cal O}(#1)$}
\newcommand{\BigOmega}[1]{$\Omega(#1)$}
\newcommand{\BigTheta}[1]{$\Theta(#1)$}
\newcommand{\Si}{{}^{28}\mathrm{Si}}
\newcommand{\Ph}{\,{}^{31}\mathrm{P}}
\newcommand{\ceiling}[1]{\left\lceil #1 \right\rceil}
\newcommand{\faM}{\lfloor \alpha M \rfloor}
\newcommand{\C}[2]{{#1 \choose #2}}
\newcommand{\comment}[1]{}
\newcommand{\ignore}[1]{}

\title{
\begin{footnotesize}
{\it To appear in the 2005 International Symposium on Microarchitecture (MICRO-38)\\}
\end{footnotesize}A Quantum Logic Array Microarchitecture: Scalable \\
Quantum Data Movement and Computation}

\author{
Tzvetan S. Metodi$^{\dag}$, Darshan D. Thaker$^{\dag}$, Andrew W.
Cross$^{\ddag}$\\Frederic T. Chong$^{\S}$ and Isaac L. Chuang$^{\ddag}$
\affiliation{{\normalsize $^{\dag}$University Of California at Davis, \textsl{\{tsmetodiev, ddthaker\}@ucdavis.edu}}\\
{\normalsize $^{\S}$University Of California at Santa Barbara, \textsl{chong@cs.ucsb.edu}}\\
{\normalsize $^{\ddag}$Massachusetts Institute of Technology, \textsl{\{awcross, ichuang\}@mit.edu}}}}

\maketitle

\thispagestyle{empty}

\sloppypar{

\begin{abstract}
Recent experimental advances have demonstrated technologies capable
of supporting scalable quantum computation.  A critical next step is
how to put those technologies together into a scalable,
fault-tolerant system that is also feasible.  We propose a Quantum
Logic Array (QLA) microarchitecture that forms the foundation of
such a system. The QLA focuses on the communication resources
necessary to efficiently support fault-tolerant computations.  We
leverage the extensive groundwork in quantum error correction theory
and provide analysis that shows that our system is both
asymptotically and empirically fault tolerant. Specifically, we use
the QLA to implement a hierarchical, array-based design and a
logarithmic expense quantum-teleportation communication protocol.
Our goal is to overcome the primary scalability challenges of
reliability, communication, and quantum resource distribution that
plague current proposals for large-scale quantum computing.  Our
work complements recent work by Balenseifer et al \cite{Oskin05a},
which studies the software tool chain necessary to simplify
development of quantum applications; here we focus on modeling a
full-scale optimized microarchitecture for scalable computing.
\end{abstract}

\Section{Introduction}\label{sec:intro}

Quantum computation exploits the ability for a single quantum bit, a
qubit, which can be implemented by the polarization states of a
photon or the spin of a single atom, to exist in a superposition of
the binary ``0'' and ``1'' states (simply denoted as $\alpha\ket{0}
+ \beta\ket{1}$, where $\alpha$ and $\beta$ are probability
amplitudes satisfying $|\alpha|^2 + |\beta|^2 = 1$). With $N$ qubits
a quantum computer can be in $2^N$ unique states at any given time.
These states can be inter-correlated such that a single logic gate
can act on all possible $2^N$ states. The exponential speedup
offered by quantum computing, based on the ability to process
quantum information through gate manipulation \cite{Deutsch85}, has
led to several quantum algorithms with substantial advantages over
known algorithms with traditional computation. The most significant
is Shor's algorithm for factoring the product of two large primes in
polynomial time. Additional algorithms include Grover's fast
database search~\cite{Grover96}; adiabatic solution of optimization
problems~\cite{Childs2002}; precise clock synchronization
\cite{Chuang2000b}; quantum key distribution~\cite{BB84}; and
recently, Gauss sums~\cite{vandam02a} and Pell's
equation~\cite{Hallgren02a2}.

A relevant large-scale quantum system must be capable of reaching a
system size of $S = KQ \geq 10^{12}$, where $K$ denotes the number
of computational steps and $Q$ denotes the number of computational
units. Quantum data is inherently very unstable, which leads to a
lack of reliable operations that can be performed on it. Also if
left idle, this quantum data will interact with its environment and
lose state, a process called {\it decoherence}. Finally, there is
the difficulty of transmitting quantum data between computational
units without losing state. This implies that the greatest challenge
towards a large, practically useful quantum computer, is designing
an architecture that incorporates the required amount of
fault-tolerance while minimizing overhead.

Previous work in large-scale quantum architecture
\cite{Oskin02,Oskin03,Copsey03} has led to the consideration of
several main scalability issues that must be taken into account:
reliable and realistic implementation technology; robust error
correction and fault-tolerant structures; efficient quantum resource
distribution.

\noindent{\bf 1. Reliable and realistic implementation technology:}
There are multiple approaches from very diverse fields of science
for the realization of a full-scale quantum information processor.
Solid state technologies, trapped ions, and superconducting quantum
computation are just a small number of many physical implementations
currently being studied. No matter the choice, any technology used
to implement a quantum information processor must adhere to four
main requirements~\cite{DiVincenzo00}: {\bf 1)} It must allow the
initialization of an arbitrary $n$-qubit quantum system to a known
state. {\bf 2)} A universal set of quantum operations must be
available to manipulate the initialized system and bring it to a
desired correlated state. {\bf 3)} The technology must have the
ability to reliably measure the quantum system. {\bf 4)} It must
allow much longer qubit lifetimes than the time of a quantum logic
gate. The second requirement encompasses multi-qubit operations;
thus, it implies that a quantum architecture must also allow for
sufficient and reliable communication between physical qubits.

\noindent{\bf 2. Robust error correction and fault tolerant
structures:} Due to the high volatility of quantum data, actively
stabilizing the system's state through error correction will be one
of the most vital operations through the course of a quantum
algorithm. Fault tolerance and quantum error correction constitute a
significant field of research
\cite{Shor95,Steane96,Knill97a,Gottesman96,Aharonov97a,Kitaev97}
that has produced some very powerful quantum error correcting codes
analogous, but fundamentally different from their classical
counterparts. The most important result, for our purposes, is the
Threshold Theorem \cite{Aharonov97a}, which says that an {\it
arbitrarily reliable} quantum gate can be implemented using only
{\it imperfect gates}, provided the imperfect gates have failure
probability below a certain {\it threshold}. This remarkable result
is achieved through four steps: using quantum error-correction
codes; performing all computations on encoded data; using fault
tolerant procedures; and recursively encoding until the desired
reliability is obtained. A successful architecture must be carefully
designed to minimize the overhead of recursive error correction and
be able to accommodate some of the most efficient error correcting
codes.

\noindent{\bf 3. Efficient quantum resource distribution:} The
quantum no-cloning theorem~\cite{Zurek} (i.e. the inability to copy
quantum data) prevents the ability to place quantum information on a
wire, duplicate, and transmit it to another location. Each qubit
must be physically transported from the source to the destination.
This makes each qubit a physical transmitter of quantum information,
a restriction which places great constraints on quantum data
distribution. Particularly troublesome is moving the qubits over
large distances where it must be constantly ensured the data is safe
from corruption. One method is to repeatedly error correct along the
channel at a cost of additional error correction resources. Another
solution is to use a purely quantum concept to implement a
long-range wire \cite{Oskin03}: teleportation~\cite{Bennett93a},
which has been experimentally demonstrated on a very small scale
\cite{Bouwmeester97a,Riebe04,Barrett04}. Teleportation transmits a
quantum state between two points without actually sending any
quantum data, but rather two bits of classical information for each
qubit on both ends.

\begin{figure}[htbp]
\centering {
\includegraphics[width=3.1in]{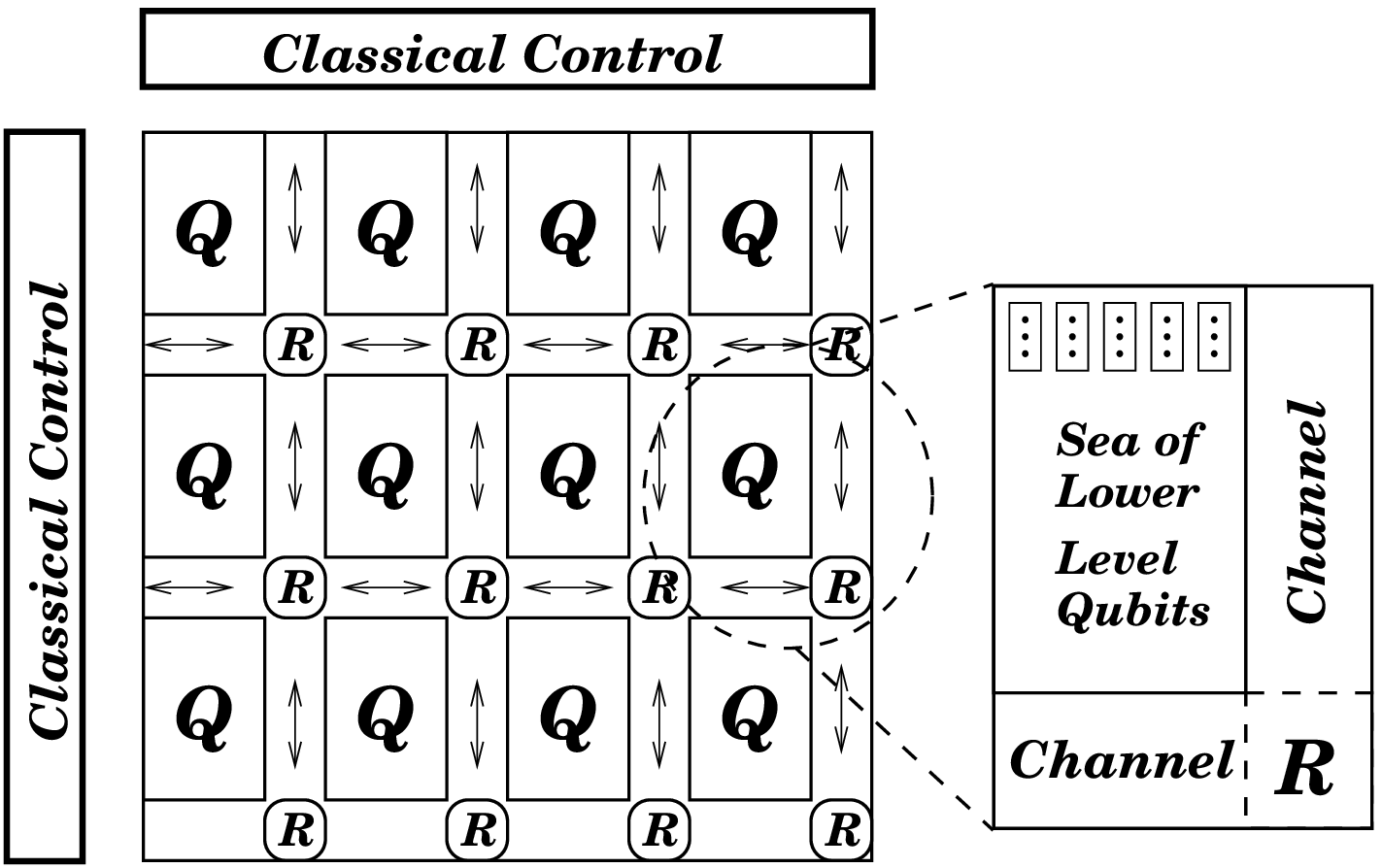}}
\caption{\textsf{High-Level quantum computer structure, where a
full-size computer consists of interconnected logical qubits
connected with programmable communication network. The letters $R$
denote an integrated switch islands for redirecting quantum data
coming from nearby logical qubits or other repeater islands.}}
\label{fig:qfpga:b}
\end{figure}

This paper introduces and evaluates the design of the Quantum Logic
Array (QLA) architecture which takes the following approach to
leveraging the three architecture requirements described above:

\vspace{0.15in}

\begin{tightenum}   

\item
At the lowest level QLA is based on the trapped ion-technology
\cite{Cirac95,Kielpinski02,Williams03a}, which uses a single trapped
atomic ion as a storage for a single unit of quantum data. In
particular QLA is based on the highly scalable model of (CCD) style
ion-trap quantum information processing architecture proposed by
Kielpinski et al \cite{Wineland98,Kielpinski02}. This model consists
of ions trapped in interconnected trap arrays and moved from trap to
trap to interact \cite{Barrett04,Riebe04}.

\item
We have designed the architecture as a block structure (Figure
\ref{fig:qfpga:b}), which fits naturally to quantum error
correction, where each building block/tile reflects the
error-correction algorithm used. QLA itself is built by tiling these
building blocks to form the hierarchies required for larger and more
reliable encodings. In addition, QLA invests area in communication
channels to allow movement of ions without hindering the parallelism
required by fault tolerant structures.

\item
By structuring the large-scale model as a datapath oriented block
architecture of independent, tightly compact computational units QLA
allows us to limit direct ion movement to shorter, local distances
within each computational unit. At larger distances (i.e. between
computational units) we employ teleportation to avoid moving data
directly over the long channels. Furthermore we couple teleportation
with the concept of quantum repeaters \cite{Dur98a} to avoid the
exponential resource overhead.

\end{tightenum} 

\noindent{\bf The Contributions of this Paper are:}~~{\bf 1)} We
propose the QLA micro-architecture, which is designed for efficient
quantum error-correction and error-free long range communication of
quantum states. {\bf 2)} While teleportation has been proposed as a
means of communication, we show the limitations of a simplistic
approach using teleportation. We then show how the QLA
micro-architecture can be effectively used to overcome these
limitations. {\bf 3)} To model QLA, we developed ARQ: a scalable
quantum architectures simulator that maps quantum applications to
fault-tolerant layouts for simulation. ARQ's input is based on the
circuit model \cite{VBE95} of quantum computation, which is the most
common representation of quantum applications, and allows the tight
integration of algorithms and architecture. The complexity of
simulating a complete $n$-qubit quantum system grows as $O(2^n)$ on
a classical machine. ARQ avoids exponential simulation costs by
simulating only a subset of the possible quantum gates, which can be
simulated in polynomial time using a mathematical stabilizer
formalism - the same formalism at the core of the most efficient
quantum error correction codes \cite{Gottesman98h,Aaronson04}. {\bf
4)} To demonstrate the utility of the QLA, we analytically evaluate
its performance when factoring a $128$-bit number using Shor's
algorithm. We have developed a scheduler to manage the communication
issues, using which we determine the bandwidth required to minimize
communication overhead. Finally, we show that the QLA, if it were to
be implemented using best foreseen ion trap technologies, might
allow the implementation of Shor's algorithm to factor a $128$-bit
number in a time on the order of tens of hours, which is
significantly faster than current classical computers might achieve.

Our work complements recent work by Balensiefer in \cite{Oskin05a},
which describes a software tool-chain for ion-trap architectures.
However, our focus is on developing a more optimized
microarchitecture based upon a more analytic approach, verified
through low-level physical simulation.  As we shall see, quantum
error-correction is a recursive process and low-level simulation is
important to account for small factors that accumulate
exponentially. Our QLA microarchitecture enables substantial
performance improvements critical to supporting full-scale
applications such as Shor's factorization algorithm \cite{Shor94}.
Balensiefer's work provides the software infrastructure to simplify
development of such applications.

The rest of this paper is organized as follows:
Section~\ref{sec:technology} gives a brief overview of trapped ion
technology. Section~\ref{sec:qfpga} then introduces the QLA
micro-processor, followed by its detailed structure and
characteristics of its components in Section~\ref{sec:structure}.
Section~\ref{sec:shor} is our evaluation of a large system executing
Shor's algorithm. Finally, we offer discussion of future work and
conclusions in Sections \ref{sec:future_work} and
\ref{sec:conclude}.

\Section{Technology Description}\label{sec:technology}

Quantum computers are no longer a fantasy for the future. In
particular, quantum ion-trapping technology may potentially lead to
a quantum computer with memory size of $50$-$100$ qubits within the
next $5$ years \cite{ARDA}. While this may seem to be a very small
computer, efforts are underway to construct prototypes that will
demonstrate the microarchitectural building blocks for a large-scale
machine.

\begin{figure}
\centering {
\includegraphics[width=2in]{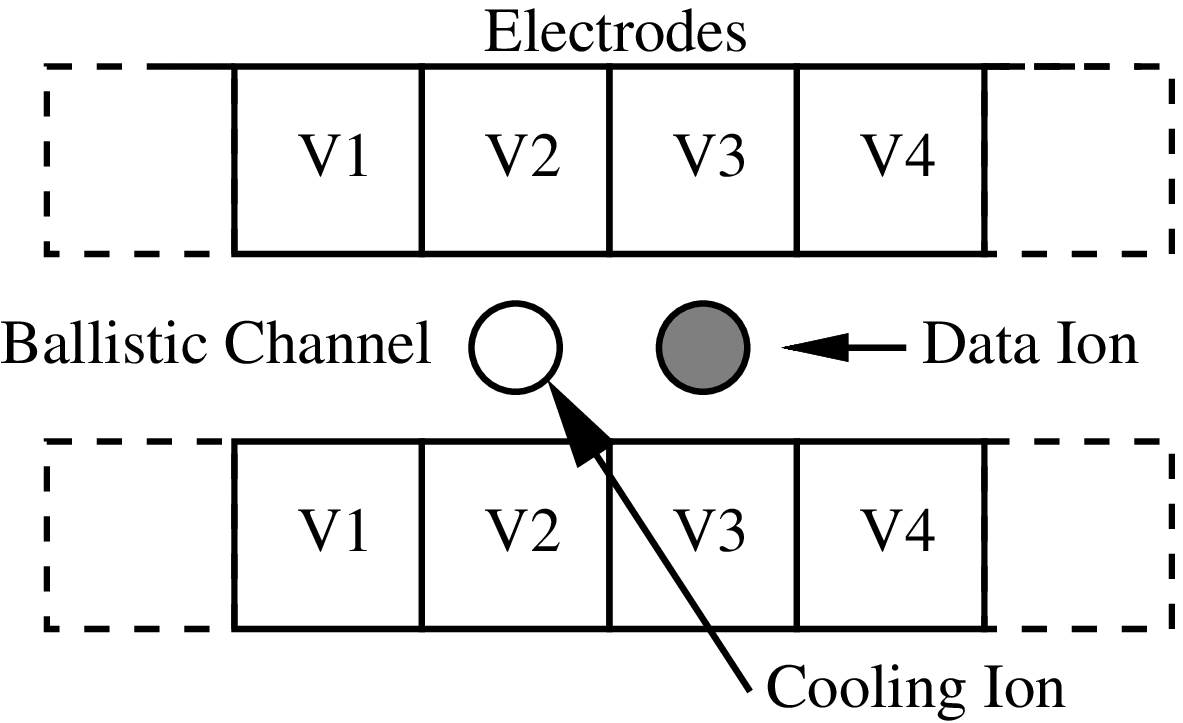}}
\caption{\textsf{A simple schematic of the basic elements necessary
for trapped ion quantum computing.}}\label{fig:ion_model:a}
\end{figure}

A high-level schematic of our ion-trap quantum information processor
is shown in Figure \ref{fig:qfpga:b}. The figure shows a number of
logical computational units (denoted by the letter $Q$) separated by
long range teleportation based communication channels. Each
computational unit is a sea of physical atomic ions as shown in
Figure \ref{fig:ion_model:a}. The quantum ion-trap processor is
surrounded by classical processors, which are used to control the
execution of almost everything, from processing quantum measurement
information to scheduling of the laser pulses that operate on the
ions.

We now take a step back and give a brief overview of the ion-trap
technology followed by the expected technology parameters in
Subsection \ref{sec:params}.

\SubSection{Ion-Traps: a Brief Overview}

Ion-trap quantum computation, initially proposed by Cirac and Zoller
~\cite{Cirac95}, uses  a number of atomic ions that interact with
lasers to quantum compute. Qubits are stored in the internal
electronic and nuclear states of the ions and the traps themselves
are segmented RF Paul traps that allow individual ion addressing
(Figure \ref{fig:ion_model:a}). Two ions in neighboring traps can
couple to each other forming a linear chain. The vibrational modes
of this chain allow a number ions to interact for multi-qubit
quantum gates, which together with single qubit rotations yield a
universal set of quantum logic. All quantum logic, including
measurement, is implemented by applying lasers on the target ions.
Individual ions are measured through state-dependent resonance
fluorescent readout, where $\ket{1}$ fluoresces weakly and $\ket{0}$
very strongly \cite{Hahn50}.

\begin{figure}[htbp]
\centering {
\includegraphics[width=1.7in]{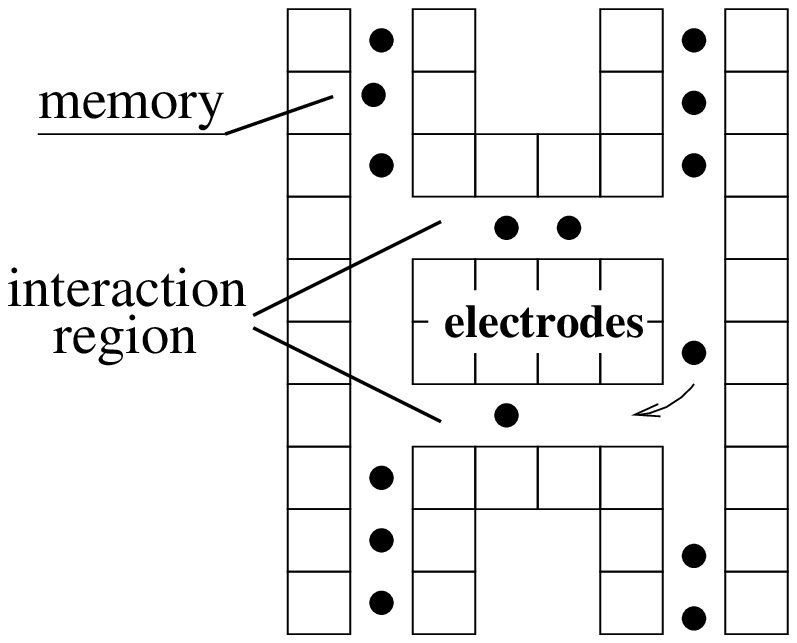}}
\caption{\textsf{The $QCCD$ model proposed by Kielpinski, Monroe,
and Wineland in \cite{Kielpinski02}. Ions are {\em ballistically}
shuttled from region to region by changing the trapping voltage
potentials.}}\label{fig:ion_model:b}
\end{figure}

When ions are manipulated they acquire vibrational heating, which
has a negative effect on the gate fidelities. To avoid direct
application of cooling lasers on the data ions, which would destroy
the quantum information, sympathetic recooling ions are used (as
shown in Figure \ref{fig:ion_model:a}), which are always kept at a
cooled ground state and are used to absorb much of the vibrational
energy from the data ions.

This original proposal, however, does not scale well \cite{Rowe02}.
As the linear chain size increases the vibrational modes of the ions
become harder and harder to identify, thus reducing gate fidelities.
Kielpinski, Wineland, and Monroe at NIST have proposed a scalable
microarchitecture using interconnected linear trap arrays
\cite{Kielpinski02}. Using multiple traps allows for greater control
over the logic gates by reducing the size of the linear ion chains.
The physical location of each ion within the network is defined by
the externally adjustable trapping voltages of the electrodes around
the ion. By changing the neighboring voltage potentials ions can be
{\em ballistically} moved from trap to trap, thus allowing the
system to handle a very large number of qubits (or ions). This type
of ion-trap network is called a Quantum Charge Coupled Device, or
simply a $QCCD$ (see Figure \ref{fig:ion_model:b}), and has been
realized with current alumina micromachining techniques.

Our abstraction of the $QCCD$ model assumes that the QLA
microarchitecture is a $2$-D grid of identical cells, where all
cells are attached on the alumina substrate. Cells can contain an
ion, electrode, or just be empty to allow a {\em ballistic channel}
for shuttling ions around as shown in Figures \ref{fig:ion_model:b}
or \ref{fig:qfpga:a}. We do not make distinction between memory and
interaction regions as in the original proposal, but allow quantum
logic, along with qubit initialization to be performed anywhere in
the layout. This allows the reuse of ions as the algorithm
progresses.

\noindent\textbf{Ballistic Channels Latency and Bandwidth:} Previous
work~\cite{Oskin03} has studied in detail quantum channels which
consist of swapping the information from qubit to qubit. The
ion-trap case is equivalent if we think of the information being
moved on an ion cell by cell along a channel of empty cells. The
latency is proportional to the number of cells traversed. If $D$ is
the number of cells and $T$ is the time to go from cell to cell,
then the total time of the trip is $(\tau + (T \times D))$, where
the split time $\tau=10\mu$s, is the initial cost of starting a
movement operation across a channel by splitting the ion from its
current chain. Considering a trap of $20\mu$m as suggested in
Reference \cite{Steane04b} a single trap can be traversed with a
time cost of $T=0.01\mu$s. The independence of the electrode cells
from one another allows the ions to move in parallel; thus,
pipelining a single channel. In this manner, the ballistic channels
provide a bandwidth of $\approx 100$M qbps (qubits per second).

\SubSection{Technology Parameters}\label{sec:params}

\begin{table}
    \begin{center}
    \begin{footnotesize}
    \begin{tabular}{|c|c|c|c|}
        \hline
        Operation & Time & $P_{current}$ & $P_{expected}$\\
        \hline 
                     &              &                 &                 \\
        Single Gate  &  $1\mu$s     &   $0.0001$      &   $10^{-8}$     \\
        Double Gate  &  $10\mu$s    &   $0.03$        &   $10^{-7}$     \\
        Measure      &  $100\mu$s   &   $0.01$        &   $10^{-8}$     \\
        Movement     &  $10ns/\mu$m &   $0.005/\mu$m  &   $10^{-6}/$cell\\
        Split        &  $10\mu$s  &                 & \\
        Cooling      &  $1\mu$s     &                 & \\
        Memory time  &  $10-100$ sec   &                 & \\
        \hline
    \end{tabular}
    \end{footnotesize}
    \end{center}
    \caption{\textsf{Column $1$ gives estimates for execution times for basic physical
    operations used in the QLA model.  Column $2$ gives currently achieved
    component failure rates $P_{current}$, based on experimental measurements
    at NIST with $^9Be^+$ ions, and using $^{24}Mg^+$ ions for sympathetic
    cooling \cite{Wineland98,Leibfried03}. Column $3$ gives projected component
    failure rates $P_{xpected}$, extrapolated following the ARDA quantum computation
    roadmap \cite{ARDA}, and discussions with the NIST researchers; these
    estimates are used in modeling the performance of the
    QLA design.}}\vspace{-0.3in} \label{table:ion_params}
\end{table}

Table~\ref{table:ion_params} shows a summary of the physical
parameters used in our QLA architecture, to model the performance of
ion-trap computation. The current experimentally achieved component
failure rates are denoted as $P_{current}$, while the expected
failure rates, $P_{expected}$, are based on {\em best-possible}
experimental implementations for the technology motivated by recent
ion-trap literature \cite{ARDA,Steane04b,Ozeri05}. The parameters
are justified by the fact that the current challenges with the
ion-trap technology are technical; current issues include electrode
surface integrity, the structure of the substrate, and precise
control of the laser phase, polarization, spatial delivery, and
timing stability. Movement errors could be substantially reduced by
improving the trap electrode surface integrity \cite{Rowe02}. The
quality of the trap surface also directly affects movement and gate
speed, since its improvement should substantially reduce motional
heating. Using semiconductor materials for the trap implementation
has been proposed in \cite{Kielpinski02}; this is a technique which
should significantly improve the electrode surfaces. Furthermore,
precise control of the laser parameters as described in Reference
\cite{Ozeri05} can significantly improve the reliability of the
quantum logic gates.

Anticipating advances in ion trap technology and techniques, we
choose space and timing parameters for the QLA design as follows. We
let the trap separation be $\approx 20\mu$m. Turning a corner at
$QCCD$ channel intersections is a complicated operation that adds
additional motional heating on the ion-qubit.  We will let
corner-turning speed be equivalent to the time for splitting two
ions from a linear chain of $10\mu$s. In addition QLA is designed in
such a way that no single gate will require more than two turns when
we are using direct ballistic communication, and no turns at all
when we are using teleportation.

\Section{The QLA Architecture}\label{sec:qfpga}

This section provides a brief overview of the QLA architecture
(Figure \ref{fig:qfpga:b}). The intent is to introduce the reader to
the high level structure of the system. The component details and
our low level design decisions are left for Section
\ref{sec:structure}, which follows next.

\noindent\textbf{Block Structure for Error Correction:} The
underlying structure of QLA is intended for error correction, by far
the most dominant and basic operation in a quantum machine
\cite{Oskin02}. Error correction is expensive because arbitrary
reliability is achieved by recursively encoding our qubits at the
cost of both exponential resource and operations overhead. Recursive
error correction works by encoding $N$ physical ion-qubits into a
known highly correlated state that can be used to represent a single
logical data bit. This data bit is now at level $1$ recursion and
will have the property of being in a superposition of ``0'' and
``1'' much like a single physical qubit. Encoding once more we can
create a {\em logical qubit} at level $2$ recursion with $N^2$
physical ion-qubits. With each level, $L$, of encoding the
probability of failure of the system scales as $p_{0}^{2^L}$ as we
will see in Section \ref{sec:structure}, where $p_0$ is the failure
rate of the individual physical components.

\begin{figure}[htbp]
\centering {
\includegraphics[width=1.8in]{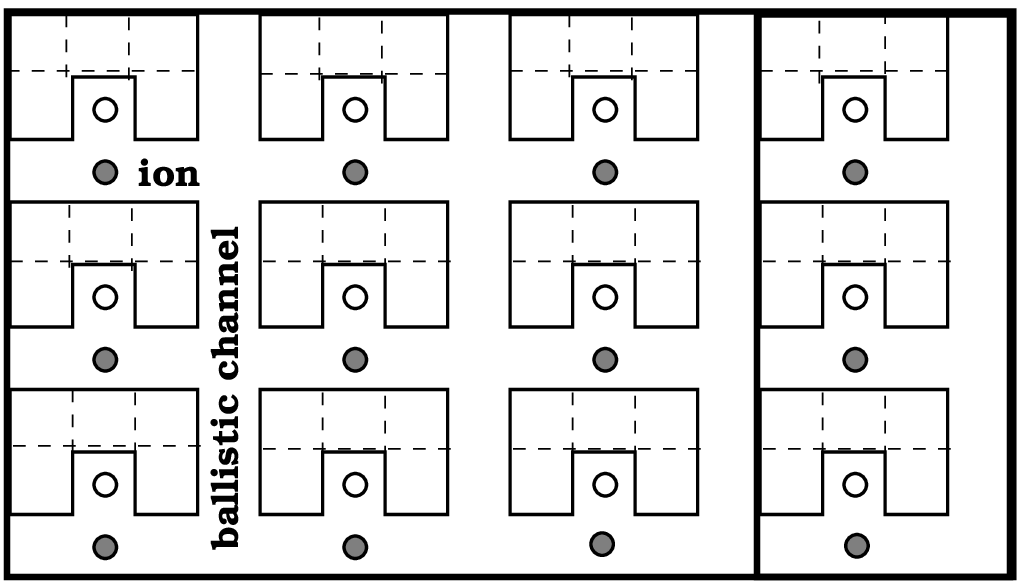}}
\caption{\textsf{The building blocks of the QLA microarchitecture.
Shown are 4 Level 1 building blocks, where the far right side
outlines a single block. The circles are data ions (solid) and
sympathetic cooling ions clear).}}\vspace{-0.1in}\label{fig:qfpga:a}
\end{figure}

The QLA structure fits naturally to quantum error correction because
the structure of the building blocks reflects the error-correction
algorithm used. Each basic building block represents a single
level-one logical qubit as shown in Figure \ref{fig:qfpga:a}. For
simplicity, Figure \ref{fig:qfpga:a} is drawn to show the level $1$
blocks of a $3$-bit error correcting code, but the structure is
easily extended to $7$-bit and larger codes \cite{Steane02a}. As the
figure shows, each building block consists of data ions supported by
their cooling ions and trapped between the electrode cells. The
investment in communication channels for ballistic ion movement
around the physical qubits allows us to limit the high costs of
turning. Any two qubit gate at any level of encoding requires at
most $2$ turns per physical ion in each direction. Furthermore, the
adaptability of the QLA design to the application being executed
allows us to structure the logical qubits such that they fully
comply with the fault-tolerant error correction requirements in
References~\cite{Gottesman99} and \cite{Svore04a}, which demand
utilizing maximum parallelism and locality. We empirically
demonstrate the fault tolerant property of our design in Section
\ref{sec:structure}.

\noindent\textbf{Logical Interconnect:} The computational units
denoted by the letter $Q$ in Figure \ref{fig:qfpga:b} are in fact
encoded logical qubits that represent a single qubit of information
whose detailed implementation is described in Section
\ref{sec:qubit}. Each logical qubit is a regular structure of
physical ions as shown in Figure \ref{fig:qfpga:a} controlled by
sequences of {\em laser pulses}. The logical qubits are positioned
on the substrate in a regular array fashion, connected with a
tightly integrated repeater-based \cite{Dur98a}  interconnect as
shown in Figure \ref{fig:qfpga:b}. This makes the high-level design
very similar to classical tile based architectures. The key
difference is that the communication paths must account for data
errors in addition to latency. The communication paths are composed
of similar physical building blocks as the logical qubits. The
integrated repeaters denoted with the letter $R$ in Figure
\ref{fig:qfpga:b} are called {\em teleportation islands} that
redirect traffic in the $4$ cardinal directions by teleporting data
from one repeater to the next. As we will see in Section
\ref{sec:shor}, this interconnect design is one of the key
innovative features of our quantum architecture, as it allows us to
completely overlap communication and computation, thus eliminating
communication latency at the application level of the program.

\noindent\textbf{Programming The Architecture:} All scheduling and
physical control is performed by the classical processors
surrounding the quantum machine. Since physical quantum operations
have a latency several orders of magnitude larger than classical
operations, a sophisticated classical processor will easily be able
to schedule the operations at run-time throughout the execution of
the algorithm.

Our general purpose quantum simulator ARQ takes a description of a
general quantum circuit with a sequence of quantum gates as an
input, maps it onto a specified physical layout, and generates pulse
sequence files, which are then executed on the general quantum
architecture simulator. For scalability, an actual ion-trap system
could manipulate qubits by focusing a small number of lasers through
a MEMS mirror array as used in optical routers \cite{MEMS-mirror}.
The optimization of our algorithms to use the smallest number of
lasers, essentially making them more effective for SIMD
architectures, is a subject of future research. A tool chain to
generate such optimized schedules is also an open area. Our focus is
the design of the microarchitecture and its evaluation through
hand-optimized applications.

\Section{Components of the Architecture}\label{sec:structure}

This Section describes in more detail the different components of
our architecture, along with the design decisions and assumptions we
have made in the process of developing QLA. First we describe each
logical qubit (Section \ref{sec:qubit}), which is followed by a
description of the logical interconnect (Section
\ref{sec:transport}).

Although the analysis in the following sections becomes somewhat
detailed, the key concept is that the structure of QLA supports
arbitrary quantum gates such that reliability is increased. We
empirically verify the fault tolerant structure of our logical qubit
in Subsection \ref{sec:numerical_analysis}; however, at this stage
of the design we cannot rely on simulation alone. We use the
simulation to validate the analytical intuition that forms the basis
of our qubit. We cannot generalize the data to other designs for two
reasons: \textbf{1)} Data may have multiple inflection points
\cite{Svore05a} and we might be misled by the analysis of just one
point. \textbf{2)} We find that level $2$ recursion is sufficient,
however, it is hard to empirically predict the behavior of a system
encoded at higher levels.

\SubSection{The Logical Qubit Design}\label{sec:qubit}

The logical qubit design we present is driven by the expected
ion-trap parameters (see Table \ref{table:ion_params}, column $4$),
which place us far below the error threshold required by the
threshold theorem, and allow us to optimize for both time and space.
Particularly important is the fact that the lifetime of an ion
($\approx 10$ sec) is far larger than quantum operations which are
on the order of tens of microseconds. The relatively low memory
error rates allow us to significantly reduce the area of a logical
qubit by reducing the parallelism within a single error correction
cycle, and the ancillary qubits required by the error correction
algorithm.

\begin{figure}[htbp]
\centering {
\includegraphics[width=2.5in]{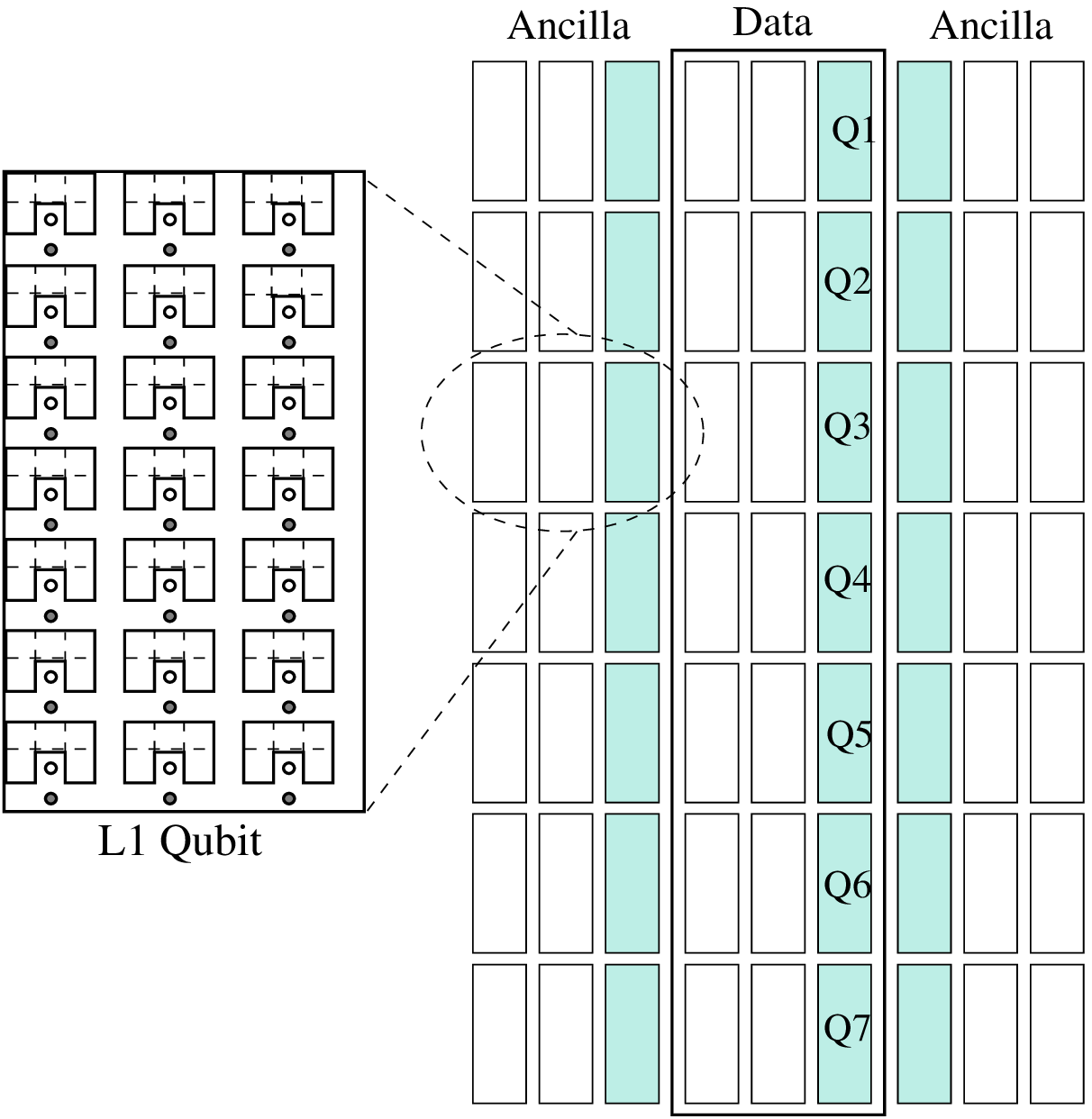}}
\caption{\textsf{The Logical Qubit: $7$ groups of $3$ level $1$
blocks make a single level $2$ logical qubit (middle). The two
identical conglomerations on the sides are ancillary blocks used for
error correction. The shaded boxes of the level $2$ qubit are the
encoded data level $1$ blocks, which are supported by their
respective level $1$ ancilla
blocks.}}\vspace{-0.1in}\label{fig:L2_qubit:a}
\end{figure}

Figure~\ref{fig:L2_qubit:a} shows the full implementation of a level
$2$ qubit. One of the most important design decisions we have made
for each logical qubit, is that it must be a self-contained unit
that requires no external {\em quantum} resources to perform logical
gates and state stabilization (i.e. error correction). This will
allow an application level compiler to divide the quantum program
into distinct data independent threads that are executed on separate
computational units, which are simply the logical qubits.

Another important design choice is the error correction code,
because it will directly dictate the amount of time each operation
requires and the size of the qubit. We choose to model the Steane
\ecc code, where $7$ physical qubits are encoded to form $1$ logical
qubit that can correct at most $(3-1)/2 = 1$ error. Our choice of
the \ecc code means that a single data logical qubit at level $2$ is
built by stacking $7$ level $1$ blocks. However, each level $1$
block must be error corrected at level $1$, so to each one we attach
two more blocks used as ancilla. To add level $2$ error correction
we add two more identical ancilla structures at level $2$ on both
sides of the data logical block. The result is
Figure~\ref{fig:L2_qubit:a}. We choose the \ecc code because it
allows the implementation of a universal set of logical gates {\em
transversally}. This means that a logical quantum bit-flip gate on
our qubit can be implemented by applying $49$ physical bit-flip
gates on the ions, in parallel.


\SubSubSection{Error Correction Latency of our Qubit}

Here we estimate the time required for each error correction step at
level $2$ recursion assuming the expected ion-trap parameters from
Table~\ref{table:ion_params}. We find that the time to complete a
single error correction step at levels $1$ and $2$ is approximately
$0.003$ and $0.043$ seconds respectively. In our design of the
logical qubit we have taken advantage of the low memory failure rate
of physical ions to minimize the physical ancilla required at the
expense of added error correction time.

The latency times were determined by analyzing the circuit shown in
Figure \ref{fig:L2_qubit:b},which demonstrates the \ecc error
correction procedure. In this representation time goes from left to
right and various one and two-qubit gates act on each line of
qubits. Each line in the circuit denotes a single encoded logical
qubit at level $2$, and at level $1$ in the lower preparation stage.

\begin{figure}
\centering {
\includegraphics[width=3.3in]{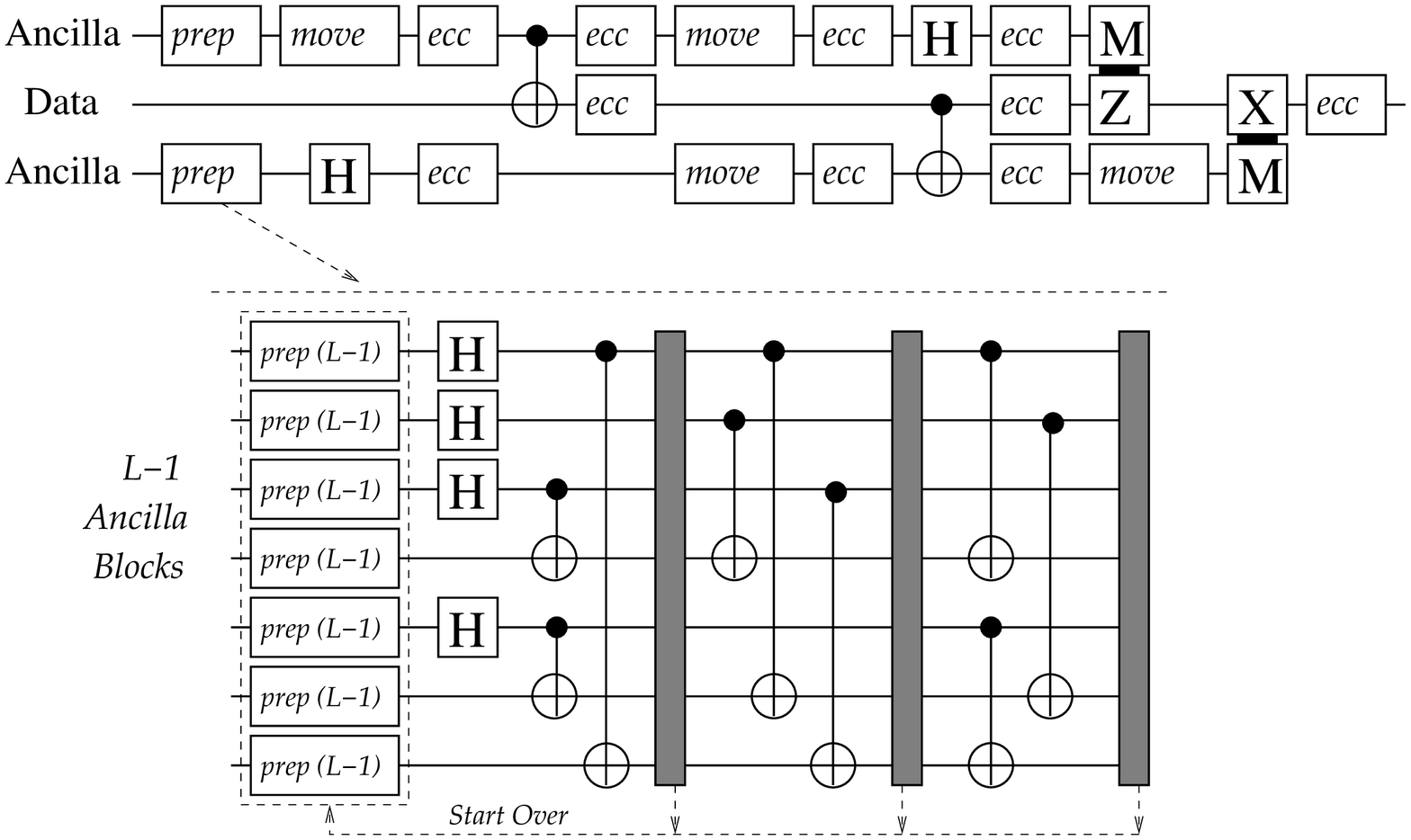}}
\caption{\textsf{Steane \ecc error correction circuit at level $L$
encoding. The top portion is the circuit with one level $L$ data
block and two identical ancilla blocks. The boxes represent logical
gates or sequences of gates. The \texttt{prep} boxes are ancilla
preparation, \texttt{move} is movement from one block to the next,
and \texttt{ecc} is error correction of that logical qubit. The
bottom portion of the circuit is a zoomed in ancilla preparation
stage from \cite{Reichardt04}. The long shaded boxes are the
syndrome extraction for each sub-logical qubit. Movement is implicit
in the $CNOT$ gates.}}\vspace{-0.1in}\label{fig:L2_qubit:b}
\end{figure}

\noindent\textbf{Computing the Latency:} The \ecc error correction
algorithm \cite{Steane97a,Steane02c,Reichardt04} consists of
extracting a syndrome to determine the location of bit-flip ($X$)
and phase-flip ($Z$) errors and applying a correction operation
based on the extracted syndrome. For each type of error, the
syndrome extraction process consists of independently encoding a
block of ancilla at the same level as the data qubit and interacting
the ancilla and the data. Clearly, the number of ancilla blocks we
have affects the parallelism we can explore when extracting
syndromes for the two types of error. For example, the level $1$
qubit shown on the left of \ref{fig:L2_qubit:a} uses $7$ ions as
data and $7$ ions as ancilla, the other $7$ are used as verification
bits of the encoding. Thus we must extract the two syndromes one
after the other. At level $2$ however we have ancilla
conglomerations on both sides of the data block (see Figure
\ref{fig:L2_qubit:a}) and we can prepare the ancilla blocks in
parallel and extract the syndromes in parallel as shown in the
circuit of Figure \ref{fig:L2_qubit:b}.

The \ecc error correction circuit in Figure \ref{fig:L2_qubit:b}
starts with syndrome extraction, which begins with the preparation
of the ancilla qubits and ends with the two measurement gates. If a
syndrome extraction yields a non-trivial syndrome (i.e. error
exists) we repeat the process until we reach two successive agreeing
error syndromes. The next step is to correct the error with the
appropriate gate followed by a lower level error correction cycle.
Equation \ref{eqn:L2_ecc_time} below, is our estimate for the error
correction latency at level $L$ recursion. We have made the
following assumptions: {\bf (a)} Two syndromes are extracted in {\em
serial} for both $X$ and $Z$ errors. {\bf (b)} We assume that in the
case of a non-trivial syndrome the next extracted syndrome will
match it, thus we can proceed with the error correction step. We
show this empirically further down. Since our logical qubit at level
$2$ is equipped with parallel syndrome extraction, assumption (a)
makes Equation \ref{eqn:L2_ecc_time} an overestimate of the final
latency:

\begin{equation}\label{eqn:L2_ecc_time}
   T_{L,ecc} =
   \begin{cases}
   2\times T_{L,synd},~~~\text{Trivial syndrome} & \\
   2(2T_{L,synd} + T_1 + T_{L-1,ecc}),~~\text{Non-trivial}&
   \end{cases}
\end{equation}

\noindent where $T_{L,synd}$ is the time to extract a syndrome at
level $L$, which is a function of the time to prepare the logical
ancilla block. $T_1$ denotes the time of a logical one-qubit gate,
and $T_{L-1,ecc}$ is the time for a lower level error correction
step that follows each level $L$ logical gate.

Numerical simulations of a level $2$ qubit showed that a non-trivial
syndrome was measured for level one with a rate of $3.35\times
10^{-4} \pm 0.41\times 10^{-4}$, and for level two at a rate of
$7.92 \times 10^{-4} \pm 0.81\times 10^{-4}$. Our simulations did
not yield a syndrome repetition of more than two times before the
error correction step. Thus, it is a reasonable assumption that in
the case of a non-trivial syndrome we require at most one more
syndrome extraction before we are ready to apply the correcting
gate. Taking a weighted average of the two cases in Equation
\ref{eqn:L2_ecc_time} we determine a level $2$ error correction time
of approximately $0.043$ seconds, where almost $0.008$ seconds is
spent in preparation of the logical ancilla.

\SubSubSection{Qubit Size and Recursion Level}

In this subsection we explain why level $2$ recursion is sufficient.
The level of recursion for each logical qubit is the most crucial
assumption for both the performance and the size of our system,
since the amount of both computational and physical resources rises
exponentially as a function of the recursion level.

A quantum computer running an application of $S = KQ$ elementary
steps (or gates), where $K$ is the number of time-steps and $Q$ is
the number of logical qubits, will require the elementary component
failure rate to be reduced to less than $P_f = 1/S$. To evaluate the
expected component failure rate at some level or recursion $L$, we
use Gottesman's estimate for local architectures \cite{Gottesman99}
shown in Equation~\ref{eqn:ftrecurselocal} below.

\begin{equation}
    P_f = \frac{1}{cr^2r^L}(cr^2p_0)^{2^L} =
         \frac{p_{th}}{r^L}(p_{th}^{-1}p_0)^{2^L},
    \label{eqn:ftrecurselocal}
\end{equation}

\noindent where the value for $r$ is the communication distance
between level $1$ blocks which are aligned in QLA to allow $r=12$
cells on average. The threshold failure rate, $p_{th}$, for the
Steane \ecc circuit accounting for movement and gates was computed
in \cite{Svore04a} to be approximately $7.5 \times 10^{-5}$. Taking
as $p_0$ the average of the expected failure probabilities given in
Table \ref{table:ion_params}, and plugging these numbers into
Equation \ref{eqn:ftrecurselocal}, we get an estimated level $2$
failure rate of $1.0 \times 10^{-16}$. This gives a computer of size
$S = KQ = 9.9 \times 10^{15}$ elementary steps. As a simple example,
we can consider Shor's factoring algorithm for a $1024$-bit number.
Employing a circuit description optimized for latency in Reference
\cite{Rod04}, we find the computer must be of approximate size $S =
4.4 \times 10^{12}$ elementary steps, which is a few orders of
magnitude below the computation size attainable with level $2$
recursion.

\SubSubSection{Numerical Analysis of the Logical Qubit}
\label{sec:numerical_analysis}

In this subsection we use ARQ to empirically compute $p_{th}$ at
level $2$ for the QLA logical qubit. Our results, displayed in
Figure \ref{fig:hadamard_plot}, show that the failure probability of
a single one-qubit logical gate rapidly drops to zero at component
failure rates lower than $p_{th} = (2.1 \pm 1.8) \times 10^{-3}$.
Above this value the rapid decrease in the reliability of our system
as recursion increases can be attributed to the additional resource
overhead of recursion.

\begin{figure}[htbp]
\centering {
\includegraphics[width=2.7in]{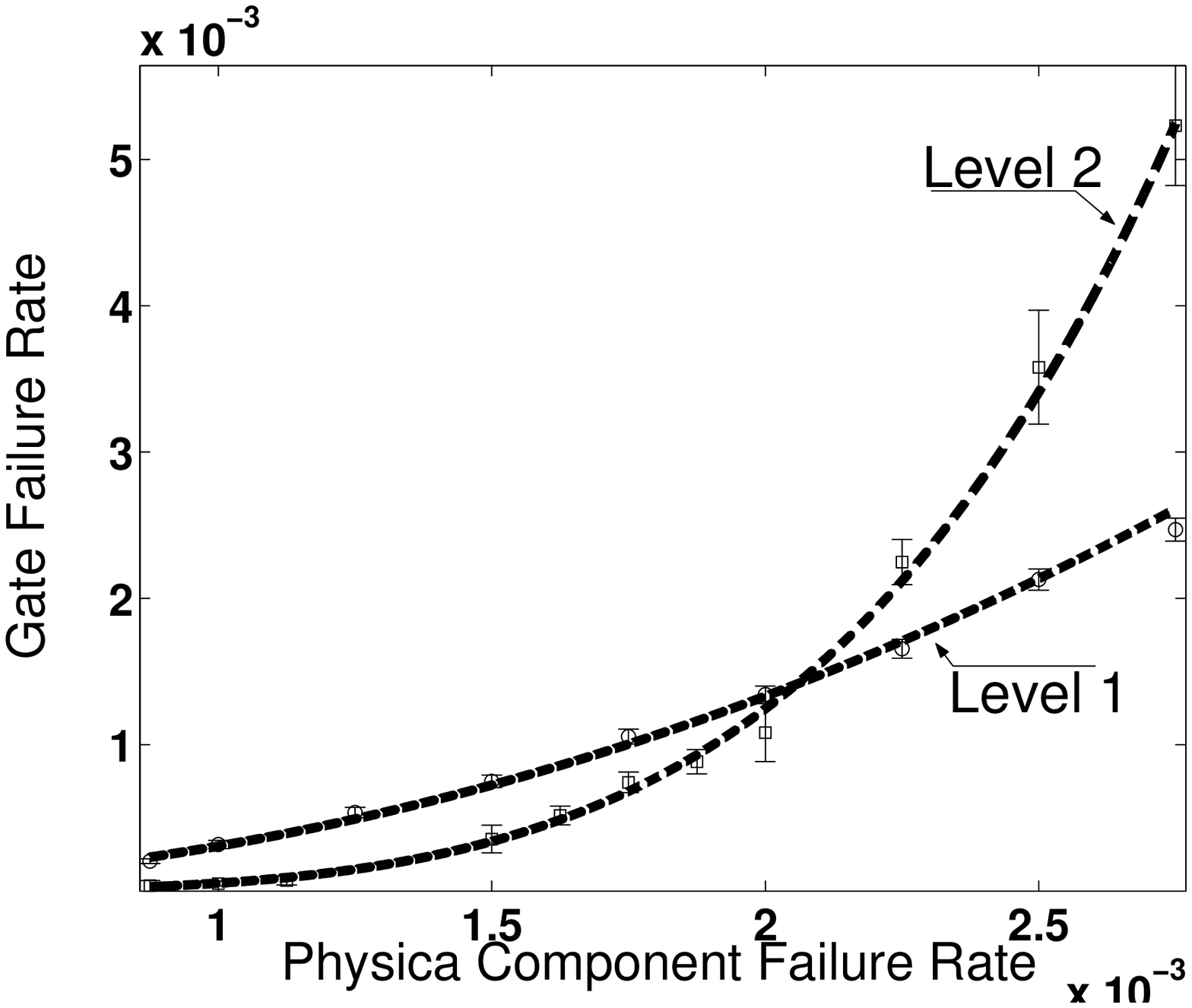}}
\caption{\textsf{Estimate of the failure probability ($\hat{y}$
axis) of a single logical one-qubit gate followed by recursive error
correction procedure at levels $1$ and $2$. The $\hat{x}$ axis
denotes individual physical component failure rates.}}
\vspace{-0.1in} \label{fig:hadamard_plot}
\end{figure}

Our estimated threshold failure probability is much higher than the
theoretical estimate of $7.5 \times 10^{-5}$ computed in
\cite{Svore04a} for several reasons; \textbf{1)} The structure of
our qubit is optimized for the error correction circuit and may vary
for different codes; \textbf{2)} The high reliability of ion-trap
memory has allowed us to significantly reduce the overall area and
ancillary resources required; \textbf{3)} The fixed, low movement
error probability, and the fact that we made the design decision to
never physically move the data, pushed our qubit's threshold closer
to the one estimated by Reichardt, $9\times 10^{-3}$, in
\cite{Reichardt04}. We observed no failure at level $2$ recursion as
the physical component errors approached the expected ion-trap
parameters from Table \ref{table:ion_params}, which was expected.
Reevaluating Equation \ref{eqn:ftrecurselocal} with the empirical
value for $p_{th}$ we get an estimated level $2$ reliability
approaching $10^{-21}$.

\noindent\textbf{Experimental Procedure:} To verify that below a
certain threshold failure rate, $p_{th}$, recursion indeed improves
the reliability of our logical qubit, we mapped the circuit in
Figure \ref{fig:L2_qubit:b} exactly to the layout shown in Figure
\ref{fig:L2_qubit:a} and simulated the execution of a single logical
one-qubit gate followed by error correction at recursion levels $1$
and $2$ respectively. As baseline technology parameters we fixed the
movement failure rate to be the expected rate shown in Table
\ref{table:ion_params}, but varied the rest of the failure
probabilities until we saw a crossing point between the two levels
of recursion.The horizontal axis of Figure \ref{fig:hadamard_plot}
marks the physical component failure probability and the vertical
axis marks the failure probability of the logical gate.

\SubSection{Logical Qubit Interconnect}\label{sec:transport}

At level $2$ recursion as described above, our qubit will have
dimensions of: $(36 \times 147)~cells = 2.11 mm^2$ at $20$ $\mu$m
large on each cell side. At this rate we can fit $100$ logical
qubits per $90nm$ technology Pentium IV processor, where each such
P4 can fit $55$ million classical transistors. As we will see in
Section \ref{sec:shor}, to factor a $1024$-bit number we may need to
communicate over a distance as large as $60$ centimeters. Given that
such a large chip can be physically realized, there are two ways to
transport quantum data at large distances while keeping it
protected: {\bf 1)} Through channels equipped for repeated error
correction of the data; and {\bf 2)} To use the concept of {\em
quantum teleportation} \cite{Bennett93a}, which requires the
exchange of classical data to recreate the state of the quantum data
in its destination. By coupling the teleportation concept with the
concept of quantum repeaters \cite{Dur98a}, we find that we can
avoid the high costs of repeated error correction and provide a
highly reliable, low-latency, fault-tolerant network interconnection
between the logical qubits. In Section \ref{sec:shor} we see that
for a high level application our network allows the complete overlap
between communication and computation. We proceed in this section
with a detailed analysis of this network.

\noindent\textbf{Quantum Teleportation:} Teleportation begins by
preparing two maximally entangled qubits $A$ and $B$ in an
Einstein-Podolsky-Rosen (EPR) state \cite{Bell64a}: $\ket{\Psi}=
\ket{0_A0_B} + \ket{1_A1_B}$. Qubit $A$ is sent to the location of
the source qubit $C$ and qubit $B$ to $C$'s intended destination.
Entangling $A$ and $C$ and measuring them allows us to recreate the
state of $C$ over the destination qubit $B$, where we have only
communicated the value of the measurement as classical data. We have
effectively teleported $C$'s state over a very large distance
without having to move it directly. As a side note, the original
states of C and A have been destroyed by the measurement, thus never
violating the no-cloning theorem.

The drawback of the teleportation scheme is that we are still
physically moving the entangled qubits $A$ and $B$; however, EPR
pairs are replaceable and with enough resources we can establish
entanglement between the source and the destination just in time for
the communication to be completed. The damaged EPR pairs can be
fixed by a process called {\em entanglement purification}
\cite{Bennett96b,Deutsch96a}, which uses ancillary EPR pairs to
distill the good ones from the bad ones. The caveat is that the
amount of resources increases exponentially with the EPR separation
distance, along with the fact that if the EPR pair becomes too
corrupted it may not even lend itself to purification.

\begin{figure}[htbp]
\centering {
\includegraphics[width=3.1in]{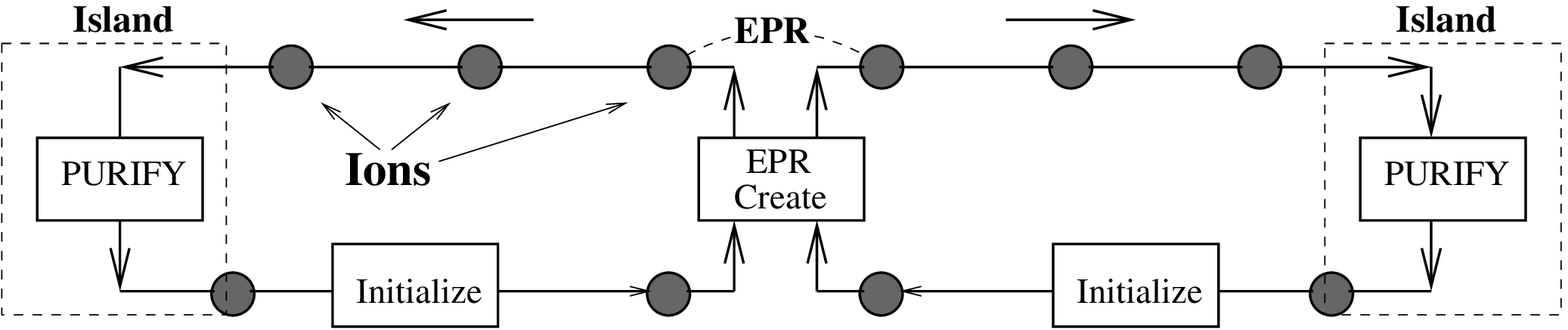}
} \caption{\textsf{Detail of a channel between two repeater
stations. The channel is two-way ballistic transport region, where
the EPR pairs are created in the middle and distributed in a
pipeline fashion to the two Island/Reapeater stations.}}
\label{fig:h_channel}
\end{figure}

\noindent\textbf{Quantum Repeaters:} The EPR transport problem can
be solved by combining the concepts of {\em quantum repeaters}
\cite{Dur98a} with entanglement purification. The quantum repeaters
are islands that are strategically placed in the channels between
the logical qubits to limit the distance traveled by each EPR pair
(see Figure \ref{fig:qfpga:b}). EPR pairs only travel to two near-by
islands, where they can be efficiently purified using the
purification protocols with some additional ancillary EPR pairs. To
expand a single entangled EPR pair between the source and the
destination over the entire channel we use a logarithmic algorithm
similar to computing transitive closure. The stages of transport are
as follows: {\bf (a)} EPR pairs are created to connect each
neighboring repeater station; {\bf (b)} We teleport in parallel
across the stations to reduce the amount of connecting EPR pairs by
half at each step, but still keep the connection between the source
and the destination; {\bf (c)} Successive teleportation steps reduce
the EPR pairs by half each time, until we have a single EPR pair
connecting the source and the destination in logarithmic number of
teleportation hops. Finally we teleport the source qubit to its
desired location when a single EPR pair spans the connection
channel.

To optimize space and performance we modeled the channels between
each island as a two-way ballistic transport region (see Figure
\ref{fig:h_channel}). Each EPR pair is created in the middle and
separated to the two opposing ends. One pair is designated as the
data EPR and is continually purified in round-robin pipeline
fashion. We assume to have enough ions to handle the maximum amount
of required purification steps without having to wait for the
creation of new EPR pairs.

 \begin{figure}[htbp]
 \centering {
 \includegraphics[width=2.7in]{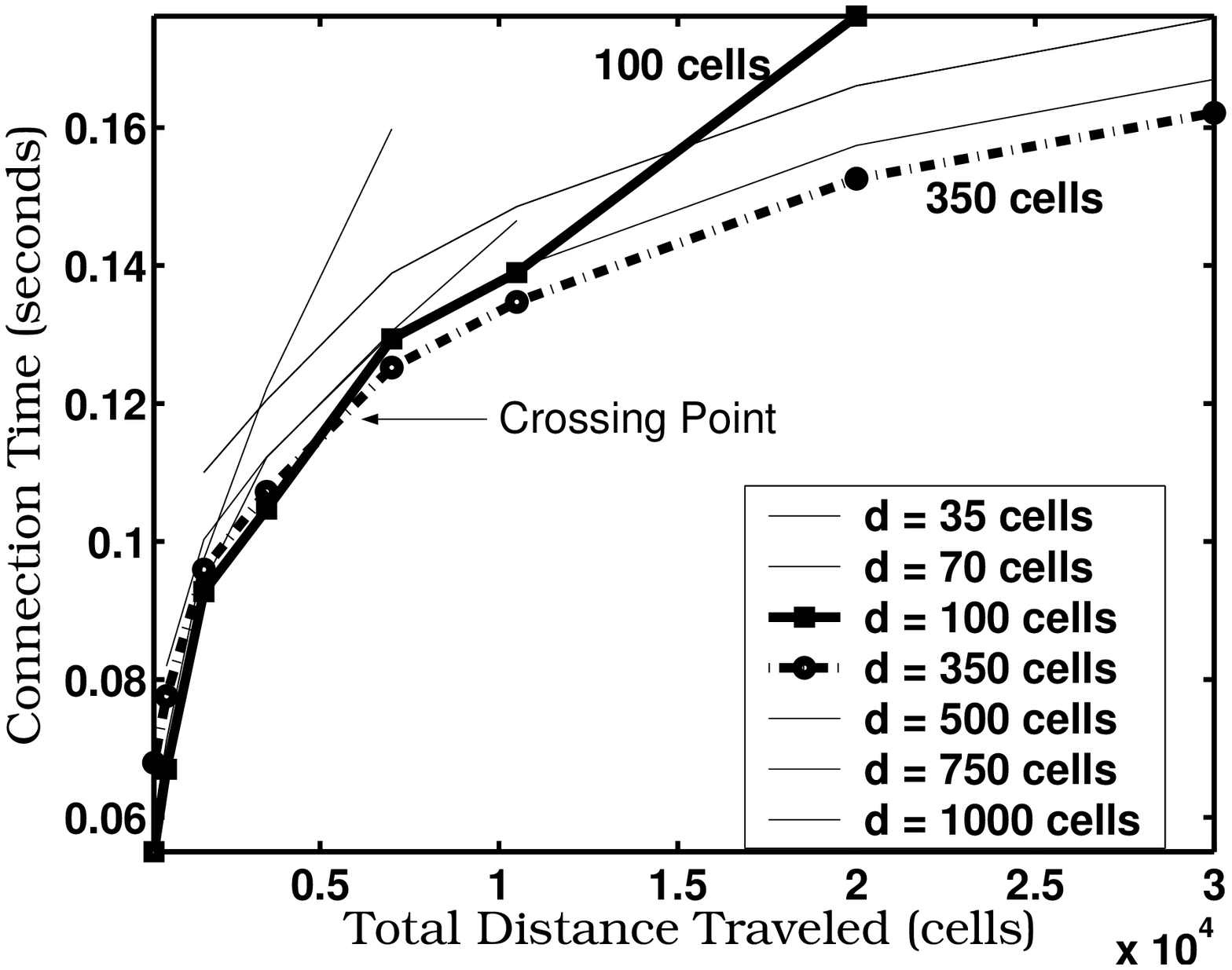}
 } \caption{\textsf{A plot of the total connection time for different
 island separation distance $d = \{35,70,100,350,500,1000\}$ cells
 between two distant qubits $A$ and $B$. With each line showing each
 distance $d$, we see that island separation of $100$ cells is more efficient
 at distances smaller than $6000$ cells (e.g. below $\approx 140$ qubits
 in the $\widehat{x}$ direction.) At larger distances separation of $350$ cells
 is preferable.}}\label{fig:separation_plot}
 \end{figure}

The teleportation islands are equipped with the capability of being
used or not being used. This allows a communication scheduler to
pick the optimal inter-island separation for the total distance
traveled. Borrowing and adapting the recursive fidelity equations
(9,19) given in \cite{Dur98a} for the Bennett purification protocol
\cite{Bennett96b}, and limiting purification to be only between two
adjacent islands we determine optimal separation between two islands
to be about $100$ cells at distances less than $\approx 6000$ cells
and about $350$ cells at greater distances (see Figure
\ref{fig:separation_plot}). In the $\widehat{x}$ direction this
amounts to an island at every third and tenth logical qubit
respectively. In the $\widehat{y}$ direction, however, we place an
island at every logical qubit due to the fact that a logical qubit
is $147$ cells in this direction. The total connection time in
Figure \ref{fig:separation_plot} was determined by adding enough
purification steps between neighboring repeater stations to avoid
purification of the final EPR pair between the source and the
destination.

\Section{QLA Performance}\label{sec:shor}

In this section we estimate the performance of QLA when executing a
general quantum application through the specific example of Shor's
factoring algorithm, which is designed to break the widely used RSA
public-key cryptosystem. RSA's security lies at the assumption that
factoring large integers is very hard, and as the RSA system and
cryptography in general have attracted much attention, so has the
factoring problem. The efforts of many researchers have made
factoring easier for numbers of any size, irrespective of the speed
of the hardware. However, factoring is still a very difficult
problem. The best classical algorithm known today \cite{Buhler94}
has complexity of:\vspace{-0.2in}

$$\exp{((1.923+o(1))({\rm log}{~N})^{1/3}({\rm
log~log}{~N})^{2/3})},$$

\noindent for an $N$-bit integer. Using this algorithm Reference
\cite{Cavallar00} has demonstrated the factorization of a $512$-bit
number in seven calendar months on $300$ fast workstations, two SGI
Origin 2000 computers, and one Cray C916 Supercomputer - a process
which amounts to $8400$ MIPS years.


Shor's quantum factoring algorithm \cite{Shor94} allows factoring of
large integers in polynomial time. The algorithm  works by using a
reduction of the factoring problem to finding the period $r$ of the
periodic function $f(x) = a^x~{\rm mod}~M$, where $a$ is a randomly
chosen number co-prime to $M$, $x$ is an integer in ${\cal
Z}_{2M^2}$, and $M$ is the number being factored. By far the
dominant part of the algorithm is this first modular exponentiation
portion, which computes $f(x)$ in superposition, over all values of
$x$.  A second part is the quantum Fourier transform (QFT), which
finds the period of $f(x)$ from the results previously computed. The
overview of the cost for factoring different $N$-bit numbers is
given in Table \ref{table:shor_numbers}. Area numbers assume scaling
along the ARDA roadmap \cite{ARDA} to $20$ $\mu$m traps, which is
also the assumed size of each cell in the QLA layout.


\begin{table}
    \begin{center}
    \begin{footnotesize}
    \begin{tabular}{|c|c|c|c|c|}
        \hline
        &&&&\\
                      & N=128  & N=512  &  N=1024   &  N=2048 \\
        \hline
        &&&&\\
        Logical Qubits & 37,971 & 150,771 & 301,251 & 602,259\\
        Toffoli Gates  & 63,729 & 397,910 & 964,919 & 2,301,767 \\
        Total Gates & 115,033 & 1,016,295 & 3,270,582 & 11,148,214\\
        Area($m^2$) & 0.11 & 0.45 & 0.90 & 1.80\\
        {\bf Time(days)} & 0.9 & 5.5 & 13.4 & 32.1\\
        \hline
    \end{tabular}
    \end{footnotesize}
    \end{center}
    \caption{\textsf{System numbers for Shor's algorithm for factoring an N-bit number
        using the circuit descriptions of \cite{Draper04, Rod04}
        and the QLA microarchitecture model. The QLA chip area is determined
        by the number of logical qubits and channels (qubits: $147 \times 36$ cells with added $11$
        and $12$ cells for the channels, where each cell is $20\mu$m
        large on each side.}}\vspace{-0.3in}
    \label{table:shor_numbers}
\end{table}

Since quantum modular exponentiation is the most computationally
intensive component of Shor's algorithm, many papers address the
need to design efficient quantum arithmetic circuits. The design of
quantum adders is specially interesting since modular exponentiation
consists of modular multiplication, which itself can be divided into
additions. We consider a quantum logarithmic-depth quantum carry
lookahead adder (QCLA) \cite{Draper04} as a component to perform
quantum modular exponentiation. The QCLA is based on ideas derived
from the classical lookahead adder. It can perform an $n$ qubit
addition with a latency of $4 \log_2 n$ Toffoli gates, 4 \cnot's and
2 \texttt{NOT}'s, and is an adder chosen from \cite{Draper04} to be
most optimized for time of computation rather than system size.

We leverage previous research \cite{Rod04} that explores various
algorithms and techniques to reduce the latency of a complete
quantum modular exponentiation.  The latency of modular
exponentiation is computed by the equation $ MExp = IM \times MAC
\times (QCLA + ArgSet) + 3p \times QCLA $, where IM is the number of
calls to the multiplier, MAC is the number of calls to the adder
block required to perform an $n$-bit modulo multiplication. ArgSet
refers to the technique of indirection which allows us to reduce the
number of multiplications. Finally, $p$ is the number of extra
qubits required by the adders for optimization, and $QCLA$ is the
depth the QCLA circuit.

We take this a step further by considering the effects of
fault-tolerance and qubit movement. As Table
\ref{table:shor_numbers} shows the dominant gate in the modular
exponentiation procedure is the Toffoli gate, which is a three qubit
controlled-controlled-\verb"NOT" gate. A fault-tolerant construction
of this gate using a universal one and two-qubit gate basis requires
$6$ additional logical ancilla qubits. The fault-tolerant Toffoli
circuit we analyze, which can be constructed following
\cite{Nielsen00a, Steane99a}, takes into consideration both the
fault-tolerant Toffoli gate and the ancilla preparation required.
The cost of a fault-tolerant Toffoli is much greater than that of
one two-qubit or a single-qubit gate. The preparation of the ancilla
qubits is an involved process of $15$ timesteps repeated three
times. However each Toffoli gate is performed on an independent set
of logical qubits; thus the ancilla preparation of each successive
Toffoli can be overlapped in most cases with the execution of the
previous Toffoli gates.

When we consider concurrency in the algorithm as a whole, it can be
seen that we can easily perform the required two qubit gates in
parallel with the Toffoli ancilla preparation. Thus, we only
consider the cost of performing fault tolerant Toffoli gates in our
overall time evaluation for the modular exponentiation. The ancilla
preparations of each Toffoli gate can be overlapped; however, in
many Toffoli's one of the three qubits involved shares its ancilla
with a previous Toffoli. Therefore each Toffoli will contribute
approximately $15$ error correction steps for the ancilla
preparation and $6$ error correction cycles to finish the gate. A
single time-step is defined by an error correction cycle since the
qubits involved at each logical gate must be error corrected each
time.

The critical component for the success of the whole design is the
cost of communication between logical qubits. We have made a design
decision that ballistic transport must be used for moving ions
within a logical qubit, and teleportation will be preferred when
moving across larger distances in order to keep the failure rate due
to movement below the threshold amount. The teleportation protocol
analyzed maintains constant cost in the face of increasing distance
and hence is a critical weapon in our armory. Since EPR pairs are
required for teleportation, we can reduce communication costs to a
minimum if we have the required number of EPR pairs available at a
logical qubit at the same time that it is ready to move.
Fortunately, this is possible because of the high cost of error
correcting the logical qubits. We can create, purify and transport
the required EPR pairs to their respective qubits while they are
undergoing error correction. But can this be done at a large scale?

To answer this question, we developed a tool to schedule the
movement of EPR pairs in QLA. We assigned one channel to carry the
created EPR pairs to their destinations and another channel to
return the used EPR pairs. Within each channel, the EPR pairs are
pipelined. We define the bandwidth of QLA's communication channels
as the number of physical channels in each direction. The distance
between each Teleportation Island was fixed at $100$ cells. The goal
of our scheduler then, is to find paths between logical qubits to
transport all the required EPR pairs within the time it takes to
perform a level $2$ error correction.


The scheduler is heuristic greedy scheduler that scalably achieves
an average of $\sim 23\%$ aggregate bandwidth utilization on our
implementation of the Toffoli gate. It works by grabbing all
available bandwidth whenever it can. However, if this means that the
scheduler cannot find the necessary paths, it will back off and
retry with a different set of start and end points. A simple
approach to doing a two qubit gate between logical qubits A and B
would be as follows: teleport A to B's physical location, perform
the gate and teleport it back. An optimization that the scheduler
incorporates is that it only moves logical qubit A back if
necessary. As a result, the logical qubits {\it drift} from one
location to another. This adds a level of complexity to the
scheduler, but at the same time reduces the amount of movement that
the qubits are subjected to.

With all the above considerations in the scheduler, we found that
given two channels in each directions (bandwidth of 2), we could
schedule communication such that it always overlapped with error
correction of the logical qubits. The end result being reliable
movement over arbitrary distances with minimal overhead.

The total time for modular exponentiation will be dominated by error
correction of the logical qubits within a fault-tolerant Toffoli
gate. For a $128$ bit number, modular exponentiation requires
$63730$ Toffoli gates with $21$ error correction steps per Toffoli.
The error correction steps of the entire algorithm amount to ($21
\times 63730$ + QFT = $1.34 \times 10^6$). Since $0.043$ seconds are
required to perform one error correction at level $2$ recursion, it
will take approximately $16$ hours to complete the factorization of
a $128$ bit number. However, assuming success of all the gates, the
circuit is repeated on average $1.3$ times \cite{Ekert96}, so the
total time to factor a $128$ bit number would be around $21$ hours.
Similar calculations lead to the execution times of the
factorization of larger integers shown in Table
\ref{table:shor_numbers}; however, the sheer sizes of the ion-trap
chips required make the physical realization of such a systems a
considerable engineering challenge, which may be impractical for
$N>128$, with current single chip technology.

\Section{Future Work}\label{sec:future_work}

The QLA architecture leverages current quantum architectures
research; however, its development must also leverage the vast
amount of knowledge and research from classical architectures.
Several critical issues quickly come to mind for the advancement of
the quantum architecture: relaxing the technology restrictions;
management of classical resources; and finally reducing the area of
the architecture.
\\\\
\noindent\textbf{Relaxing the Technology Restrictions:} Relaxing the
technology restrictions will lead to quicker realization of the QLA
microarchitecture. The assumption of the expected ion-trap failure
probabilities is not unrealistic as base parameters for a future
system such as this one. Careful simulations, however, of various
noise models and encodings must be conducted to determine how far we
can relax this assumption in order to have a system still capable of
relevant quantum computing with exponential speedup over classical
machines.

\noindent\textbf{Management of Classical Resources:} The success of
the physical implementation of the QLA model rests upon the
realization of the classical mechanisms that control the execution
of a quantum algorithm. Some of these mechanisms include; the
control of lasers for precise manipulation of thousands of logical
qubits; the amount of laser power possible; the number of
photodetectores required for measurement; and even the wiring of the
electrodes used for trapping the physical ions. Classical resources
must be optimized both in quantity and usage complexity through
clever scheduling and representation of quantum operations.
Furthermore, the inherent parallelism in quantum computation along
with the fundamentally different operations structure has the
potential to create a wide variety of interesting and very difficult
compiler design problems.

\noindent\textbf{Computer Area:} According to the results in
Table~\ref{table:shor_numbers}, the area of the ion-trap chip for
even the factoring of a $128$-bit number is roughly $0.45$ square
meters. This amounts to a chip size of $33$ centimeters at each edge
if we assume a square chip. This is a substantial fabrication and
yield challenge. QLA offers an inherent redundancy within itself,
which we can explore to raise the yield. This arises from the fact
that all logical qubits and channels are identical in both their
structure and ability to support different functionalities. Defects
can be diagnosed and masked out in software running on our classical
control processor. The fabrication challenges, however, suggest that
a multi-chip solution for solving such large problems is desirable.
Although, experimental progress has been made in this direction,
\cite{Cabrillo99,Duan01,Blinov04}, the detailed analysis of the
design and performance of such systems remains as an area for future
research.

\Section{Concluding Remarks} \label{sec:conclude}

This paper has introduced the QLA, which is a quantum computer
architecture for trapped ions, designed for efficient
error-correction and error-free communication, over arbitrary
on-chip distances. Our designs are validated by analytical reasoning
and also by simulation. We have emphasized the importance of a
datapath oriented large-scale quantum architecture for solving
realistic problems, and shown how the QLA achieves such design
goals. In addition, we have introduced ARQ, a tool used to map
quantum applications to fault-tolerant architectures.

Finally, we have shown how the QLA architecture design methodology
potentially scales, conceivably allowing a system of $7 \times 10^6$
physical ions to be able to implement Shor's algorithm to factor a
$128$-bit number within $1$ day; such performance assumes aggressive
technology parameters which are not currently achieved, but are
believed to be within reach of present experimental techniques.  The
QLA architecture should thus provide vital insight and motivation,
from a systems level perspective, to physicists involved in actually
building a large-scale quantum computer. For architects, the QLA
forms a technically sound base, that can be used to confidently
study interesting issues in quantum architectures and to work
towards more efficient, reliable, and scalable quantum computers.

\noindent {\bf Acknowledgements:} This work is supported in part by
the DARPA QUIST program, in part by a NSF CAREER grant and a UC
Davis Chancellor's Fellowship awarded to Fred Chong, and in part by
the NSF Center for Bits and Atoms at MIT

} 

\begin{footnotesize}
\bibliographystyle{spiebib}
\bibliography{micro05-paper}

\end{footnotesize}

\end{document}